\documentclass{article}
\usepackage{ijcai21}

\usepackage{times}

\usepackage{soul}
\usepackage{url}
\usepackage[hidelinks]{hyperref}
\usepackage[utf8]{inputenc}
\usepackage[small]{caption}
\usepackage{graphicx}
\usepackage{amsmath}
\usepackage{booktabs}
\usepackage{balance} % for balancing columns on the final page
\usepackage[T1]{fontenc}    % use 8-bit T1 fonts
\usepackage{hyperref}
\usepackage{wrapfig}
\usepackage{subfigure}
\usepackage{fancybox}
\usepackage{tikz}
\usepackage{caption}
\usepackage{algpseudocode}
\usepackage{algorithmicx}
\usepackage[ruled,vlined,linesnumbered]{algorithm2e}
\usepackage{diagbox}
\usepackage{multirow}
\usepackage{amsthm}
\usepackage{amsfonts}       % blackboard math symbols
\usepackage{nicefrac}       % compact symbols for 1/2, etc.
\usepackage{microtype}      % microtypography

\usepackage{graphbox}
\newcommand*{\vcenteredhbox}[1]{\begingroup
\setbox0=\hbox{#1}\parbox{\wd0}{\box0}\endgroup}

\makeatletter
\newcommand{\removelatexerror}{\let\@latex@error\@gobble}
\makeatother

\makeatletter 
  \newcommand\figcaption{\def\@captype{figure}\caption} 
  \newcommand\tabcaption{\def\@captype{table}\caption} 
\makeatother

\title{A Game-theoretic Utility Network for Cooperative Multi-Agent Decisions in Adversarial Environments}

\author{
Qin Yang\and
Ramviyas Parasuraman$^*$\\
\affiliations
 Heterogeneous Robotics Research (HeRo) Laboratory, University of Georgia \\
}

\begin{document}

\maketitle

\begin{abstract}
Underlying relationships among multi-agent systems (MAS) in hazardous scenarios can be represented as Game-theoretic models. 
We measure the performance of MAS achieving tasks from the perspective of balancing success probability and system costs.
This paper proposes a new network-based model called Game-theoretic Utility Tree (GUT), which decompose high-level strategies into executable low-level actions for cooperative MAS decisions. This is combined with a new payoff measure based on agent needs for real-time strategy games.
We present an Explore game domain to evaluate GUT against the state-of-the-art QMIX decision-making method.
Conclusive results on extensive numerical simulations indicate that GUT can organize more complex relationships among MAS cooperation, helping the group achieve challenging tasks with lower costs and higher winning rate.
\end{abstract}

\section{Introduction}

Natural systems have been the key inspirations in the design, study, and analysis of Multi-Agent Systems (MAS) \cite{wooldridge2009introduction}. \textit{Distributed Intelligence} refers to systems of entities working together to reason, plan, solve problems, think abstractly, comprehend ideas and language, and learn \cite{parker2007distributed}.
Especially for cooperative MAS, the individual is aware of other group members, and actively shares and integrates its needs, goals, actions, plans, and strategies to achieve a common goal and benefit the entire group. It can maximize global system utility and guarantee sustainable development for each group member \cite{shen2004degree}.

Systems with a wide variety of agent heterogeneity and communication abilities can be studied, and collaborative and adversarial issues also can be combined in a real-time situation \cite{stone2000multiagent}. Considering working in adversarial environments, opponents can prevent MAS from achieving global and local tasks, even impair individual or system necessary capabilities or normal functions \cite{jun2003path}.
Combining multi-agent cooperative decision-making and robotics disciplines, researchers developed the \textit{Adversarial Robotics} focusing on autonomous agents operating in adversarial environments. \cite{agmon2011multi,yehoshua2015adversarial}. 
From the robot's\footnote{Here, we use the terms agent and robot interchangeably.} needs \cite{9283249} and motivations perspective, we can classify an \textbf{Adversary} into two general categories: \textbf{Intentional} (such as enemy or intelligent opponent agent, which consciously and actively impairs the MAS needs and capabilities) and \textbf{Unintentional} (like obstacles and weather, which unaware and passively threaten MAS abilities) adversary.

\begin{figure}[t]
\centering
\begin{minipage}[b]{0.6\linewidth}
%\begin{center}
%\setlength{\abovecaptionskip}{6pt}
%\setlength{\belowcaptionskip}{-1pt}
\setlength{\abovecaptionskip}{3.5pt}
\setlength{\belowcaptionskip}{5pt}
\vcenteredhbox{\includegraphics[width=1\columnwidth]{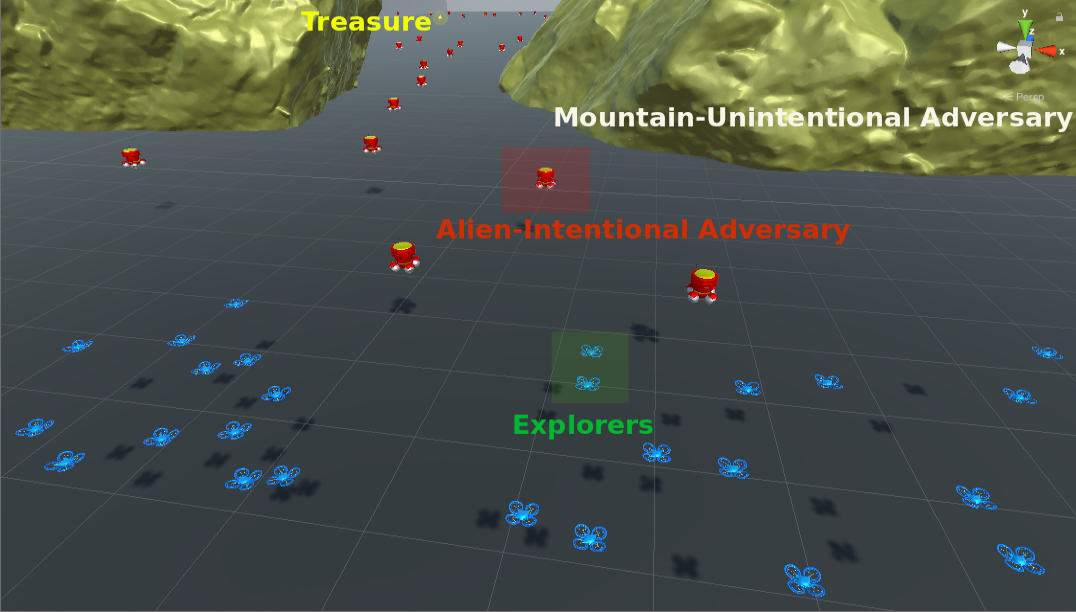}\vspace{-1.5pt}}
\caption*{\small Explore Domain}
\label{fig: explore_game}
%\end{center}
\end{minipage}
\hspace{0mm}
\begin{minipage}[b]{0.18\linewidth}
\setlength{\abovecaptionskip}{3.5pt}
\setlength{\belowcaptionskip}{3.5pt}
\includegraphics[width=1\textwidth]{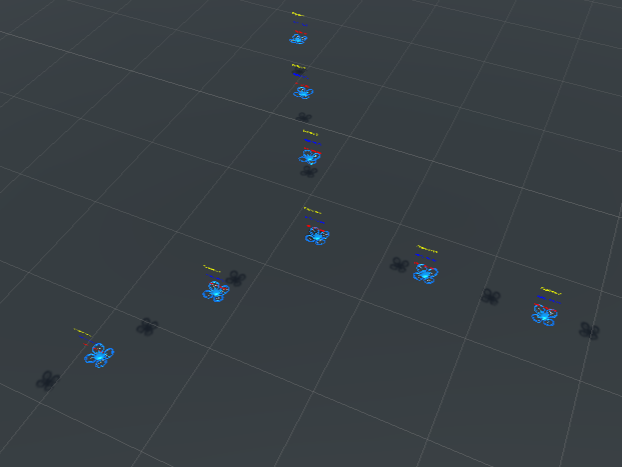}\vspace{0pt}
% \caption{\small{Patrolling}}
\caption*{\small{Patrolling}}
\label{fig: patroling_formation}
\includegraphics[width=1\textwidth]{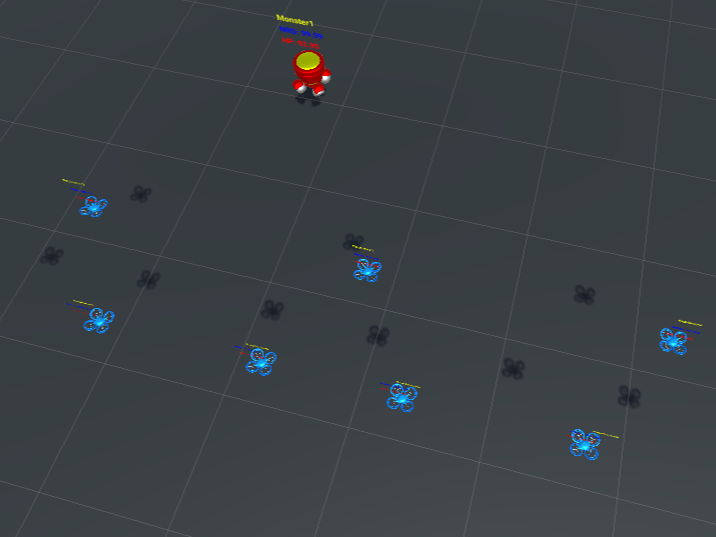}\vspace{0pt}
% \caption{\small{Attacking}}
\caption*{\small{Attacking}}
\label{fig: attacking_formation}
\end{minipage}
\begin{minipage}[b]{0.18\linewidth}
\setlength{\abovecaptionskip}{3.5pt}
\setlength{\belowcaptionskip}{3.5pt}
\includegraphics[width=1\textwidth]{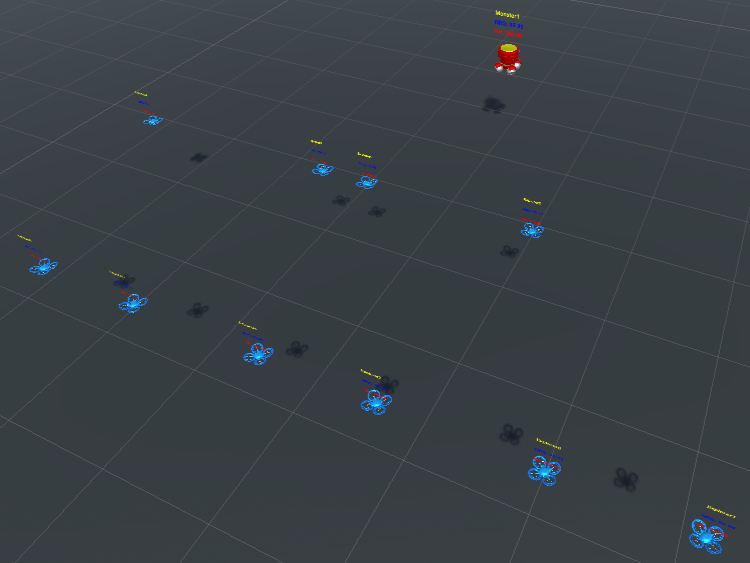}\vspace{0pt}
% \caption{\small{Defending}}
\caption*{\small{Defending}}
\label{fig: defending_formation}
\includegraphics[width=1\textwidth]{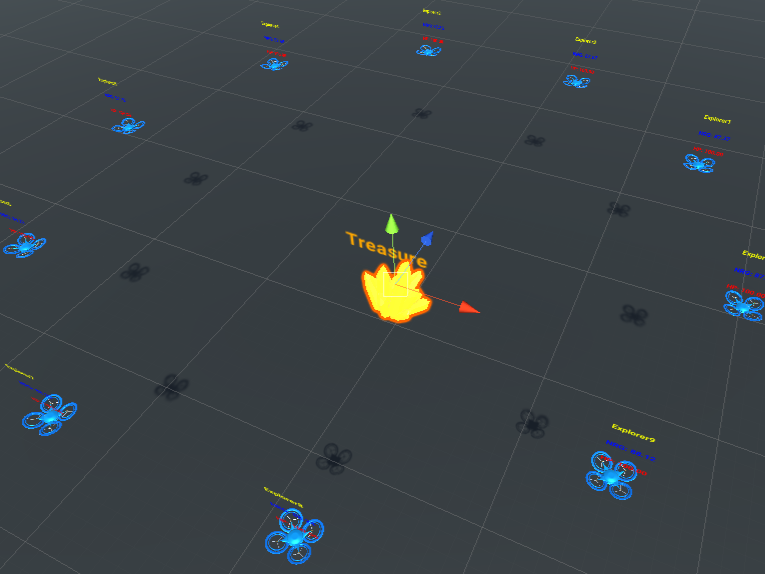}\vspace{0pt}
% \caption{\small{Circling}}
\caption*{\small{Final State}}
\label{fig: coveraging_formation}
\end{minipage}
%\vspace{-6mm}
%\setlength{\abovecaptionskip}{6pt}
% \setlength{\belowcaptionskip}{5pt}
\vspace{-4mm}
\caption{\small{An illustrative of the game scenario where Aliens block the paths to a target of Explorers}}
\vspace{-2mm}
\label{fig:overview}
\end{figure}
MAS research domains focus on solving path planning problems for avoiding static or dynamical obstacles \cite{agmon2011multi} and formation control \cite{shapira2015path,yehoshua2015adversarial} from the unintentional adversary perspective. For intentional adversaries, the "pursuit domain" \cite{Benda+JD:1986,cheng2003short} primarily deals with how to guide one or a group of pursuers to catch one or a group of moving evaders \cite{scott2017optimal,makkapati2019optimal}. Foundations for normal-form team games and extensive-form adversarial team games are provided in \cite{von1997team} and \cite{celli2018computational}, respectively.
Nevertheless, it is more realistic and practical for MAS to organize more complex relationships and behaviors, achieving given tasks with higher success probability and lower costs in adversarial environments.
%Nevertheless, it is more realistic and practical for MAS to organize more complex relationships and behaviors, achieving given tasks with higher success probability and lower costs in adversarial environments.

\paragraph{Contributions} 
This paper proposes a new hierarchical network model called \textit{Game-theoretic Utility Tree (GUT)} to achieve MAS cooperative decision-making in adversarial environments. 
%Also, from a realistic perspective, we provide a new domain -- \textit{Explorers and Aliens Game} (hereafter referred to as the "explore domain") to address the gaps discussed above.
\textit{GUT} consists of \textit{Game-theoretic Utility Computation Units} (Fig.~\ref{fig:game_decision_trees}) distributed in multiple levels by decomposing strategies, thereby significantly lowering the game-theoretic operations in strategy space dimension.
It combines the core principles of \textit{Bayesian Networks} \cite{koller2009probabilistic}, \textit{Game Theory} \cite{myerson2013game}, and \textit{Utility Theory} \cite{fishburn1970utility,kochenderfer2015decision}.
Further, we propose a novel way of calculating the payoff (utility) values through the agent needs expectations, which is also organized hierarchically similar to human needs pyramid. We also present a game of Explorers vs. Aliens (referred as \textit{"Explore domain"} - Fig.~\ref{fig:overview}) to evaluate the MAS performance from the perspective of balancing the success probability of achieving tasks and system costs by organizing involved individuals' relationships and suitable groups' strategies in adversarial environments. 
%\paragraph{Adapting The Edge} We propose a MAS collective decision algorithm called "\textit{Adapting The Edge}" (see Appendix. \ref{unintentional_adversarial_decision}) to tackle static unintentional adversaries (obstacles) integrated with GUT. 

We demonstrate the effectiveness of GUT against the state-of-the-art cooperative decision-making algorithm QMIX in extensive realistic simulations of the \textit{Explore Domain}. The results indicated that \textit{GUT} could organize more complex relationships among MAS cooperation. It helps the group achieving challenging tasks with lower costs and higher winning probability. 
The proposed approach can be applied to other Real-Time Strategy (RTS) games, which involve agents decomposing the high-level strategies into primitive actions or group atomic operations \cite{9283249} in the specific mission, such as air combat, StarCarft, robotics, etc.

\section{Background and Preliminaries}

This section briefly reviews the \textit{Bayesian Networks} and \textit{Nash Existence Theorem}, and provide a brief background to the \textit{Robot Needs Hierarchy}. We also define the adversary in adversarial environments. See Appendix~\ref{relative_definition} for other definitions.

\subsection{Bayesian Networks}

A Bayesian Network structure $\mathcal{G}$ is a directed acyclic graph whose nodes represent random variables $X_1, \cdots, X_n$. Let $Pa_{X_i}^\mathcal{G}$ denote the parents of $X_i$ in $\mathcal{G}$, and NonDescendants$_{X_i}$ denote the variables in the graph that are not descendants of $X_i$. Then $\mathcal{G}$ encodes the following set of conditional independence assumptions, called the local independencies, and denoted by $\mathcal{I}_{\ell}(\mathcal{G})$: For each variable $X_i$: $(X_i \perp$ NonDescendants$_{X_i}$$\mid Pa_{X_i}^\mathcal{G})$. In other words, the local independenceies state that each node $X_i$ is conditionally independent of its nondescendants given its parents \cite{koller2009probabilistic}.

\subsection{Nash Existence Theorem}

\textit{Nash Existence Theorem} guarantees the existence of a set of mixed strategies for finite, non-cooperative games of two or more players in which no player can improve his payoff by unilaterally changing strategy \cite{weisstein2002crc}. It guarantees that every game has at least one Nash equilibrium \cite{jiang2009tutorial}, which means that every finite game has a \textit{Pure Strategy Nash Equilibrium} or a \textit{Mixed Strategy Nash Equilibrium}.
Furthermore, in any normal-form game with constant number of strategies per player, an $\epsilon$-\textit{approximate Nash Equilibrium} can be computed in time $\mathop{O}(n^{log~n/\epsilon^2})$, where $n$ is the description size of the game \cite{daskalakis2009complexity}.

\subsection{Robot Needs Hierarchy}

In \textit{Robot Needs Hierarchy} \cite{9283249}, the agent's safety needs (Eq.~\eqref{safety_need}) expressed as the \textit{Safety Needs Expectation}, which can be calculated through its behaviors' weight and corresponding safety probability based on the data of perception and communication. The individual safety needs are the precondition for calculating the basic needs (Eq.~\eqref{basic_need}), which also can be presented as \textit{Basic Needs Expectation}. Only after fitting the safety and basic needs can consider its capability needs (Eq.~\eqref{capability_need}).
\begin{equation}
\begin{split}
    Safety~~Needs:~~ N_{s_{t_i}} = \sum_{i=1}^{s_{t_i}} W_{i} \cdot \mathbb{P}(W_{i}|P, C); \label{safety_need}
\end{split}
\end{equation}
\vspace{-4mm}
\begin{equation}
\begin{split}
    Basic~~Needs:~~ N_{b_i} = \sum_{i=1}^{b_i} W_{i} \cdot \mathbb{P}(W_{i}|P, C, N_{s_{t_i}}); \label{basic_need}
\end{split}
\end{equation}
\vspace{-4mm}
\begin{equation}
\begin{split}
    Capability~Needs: N_{c_i} = \sum_{i=1}^{c_i} A_i \cdot \mathbb{P}(A_i|T, P, N_{b_i}); \label{capability_need}
\end{split}
\end{equation}
\vspace{-4mm}
\begin{equation}
\begin{split}
    Teaming~~Needs:~~ N_{t_i} = \sum_{i=1}^{\alpha} U_i \cdot \mathbb{P(}U_i | P, C, N_{c_i}); \label{teaming_need}
\end{split}
\end{equation}

\textbf{Here,} $P$ and $C$ represent the data of agent's perception and communication separately; $T$ represents the task requirement space; $U_i$ represents the utility value of agent $i$ in the group; $W$ represents corresponding weights; $A$ represents the level of agent's corresponding capabilities based on task requirements; $b_i$, $s_{t_i}$, and $c_i$ represent the size of agent $i$'s basic, safety, and capability needs solution space respectively; $\alpha$ represents the number of agents in the group.

Through the above analysis, we adopt \textit{Utility Theory} to define the agent's fourth level needs -- \textit{Teaming Needs} (Eq. \eqref{teaming_need}), which represent higher-level needs for an intelligent agent. It can be regarded as a kind of motivation or requirements for cooperation achieving specific goals or tasks to satisfy the individual or group's certain \textit{Expected Utilities}. 
%Furthermore, for the highest (fifth) level needs -- \textit{Learning Needs}, they can help the individual achieve self-upgrade based on previous experiences and lead to the self-evolution of the whole system.
According to \textit{Robot Needs Hierarchy}, we define the adversary as follow:

\newtheorem{myDef}{Definition}
\begin{myDef}[Adversary]
For certain state $\psi_1$ $\in$ $\Psi$ and a group of agents R$_1$ given the action series a$_{1i}$ $\in$ $\Lambda$ (action space) fulfilling task T. Supposing without any interruption, the maximum teaming needs is max(N$_1$($\psi_1$, a$_{1i}$)). Considering another groups agents R$_2$ involving with reaction series a$_{2k}$. With interruption by R$_2$, group R$_1$'s need is max(N$_{12}$($\psi_{12}$, a$_{1j}$)). If Eq. \eqref{adversary} is satisfied, it can be defined R$_2$ an \textbf{Adversary} to R$_1$. In additional, if R$_2$'s corresponding expected needs with (N$_{21}$) or without (N$_2$) involving R$_1$ are not equal, then R$_2$ will be regarded as \textbf{Intentional Adversary} (Eq. \eqref{intentional}). Otherwise, we consider R$_2$ as \textbf{Unintentional Adversary} (Eq. \eqref{unintentional}).
\begin{eqnarray}
    && max(N_1(\psi_1, a_{1i})) > max(N_{12}(\psi_{12}, a_{1j})); \label{adversary} \\
    && \mathop{\mathbb{E}}(N_{21}|\psi_{21},a_{2k}) \neq N_2; \label{intentional} \\
    && \mathop{\mathbb{E}}(N_{21}|\psi_{21},a_{2k}) = N_2,~~~~i, j, k \in Z^+. \label{unintentional}
\end{eqnarray}
\end{myDef}

\section{Explore Domain Problem Statement}

In \textit{Explore Domain}, $\alpha$ Explorers are exploring and collecting rewards (reaching treasure locations $\rho_{tr}$) in an uncertain environment. Intentional ($\beta$ Aliens) and unintentional ($\gamma$ Obstacles) adversaries are randomly distributed in the scenarios.
Explorer $i$ and Alien $j$ have strategy space $S_{e_i}(s_1, ... , s_\nu)$ and $S_{a_j}(s_1, ... , s_\upsilon),~i,j,\nu,\upsilon \in Z^+$, respectively. Also, every strategy has corresponding \textit{actions} to execute $s(a_1, a_2, ... , a_r), r \in Z^+$. $C$ represents the explorers' system costs in the entire process.

Supposing Explorer's success probability (win rate) finding the treasure is $W$. We model this problem as finding a set of suitable strategies $S_e^*$ from $S_e$ under the premise of maximizing $W$ to minimize $C$ based on basic teaming needs $n_{{t_e}}$ after satisfying all the low-level needs in turn as Eq.~\eqref{teaming_need}. It can be described as an optimization problem formulating as Eq. \eqref{exploring_game_problem}.
% \vspace{-3mm}
\begin{equation}
\begin{split}
   & S_e^* = \underset{S_e}\arg [\max~~~ W(S_e | S_a, \gamma, \rho_{tr}) \\ \vspace{-3mm}
   & + \min \sum_{i=1, j=1}^{\nu, \upsilon} C_i(S_{e_i} | S_{a_j}, \gamma, N_{t_i}) ]\\ \vspace{-3mm}
   & s.t. \; \; \; N_{t_i} \geq n_{t_e},~~\forall~{i \in \alpha}.
\label{exploring_game_problem}
\end{split}
\end{equation}

\section{Approach}

Fig.~\ref{fig:game_decision_trees} outlines the structure of the \textit{Game-theoretic Utility Tree (GUT)} and its computation units distributed in each level. 
First, the \textit{game-theoretic module} (Fig.~\ref{fig:game_decision_trees} (a)) calculates the nash equilibrium based on the utility values $(u_{11}, ... , u_{nm})$ of corresponding situations, $(p_1, ... , p_{nm})$ presenting the probability of each situation.
Then, through the \textit{conditional probability(CP) module} (Fig.~\ref{fig:game_decision_trees} (b)), the CP of each situation can be described as $(p_{i1}, ... , p_{inm})$, where $p_{inm} = (p_{nm} | p_{i-1}),~i,n,m \in Z^+$. 
\textbf{Here,} $p_{i-1}$ and $S_i$ present the probability of previous situation and current Game-theoretic state; $s_a$, $s_b$ and $n$, $m$ represent their strategy space and size on both sides, respectively.
% $(s_{a1}, ... , s_{an})$ and $(s_{b1}, ... , s_{bm})$ 
% At each iteration, the agents execute the actions based on GUT decomposing high-level strategies into executable low-level behaviors. 
In this section, we explain the decision-making process in GUT and describe the specific implementation in "explore domain".

\subsection{GUT-based Decision-Making}

For intentional adversaries, agents first decompose the specific goal into several independent subtasks based on the same category of individual low-level behaviors or atomic operations (basic group strategies) \cite{9283249}. Then, through calculating various Nash equilibrium based on different situation utility values in each level's \textit{Game-theoretic Utility Computation Units}, agents can get optimal or sub-optimal strategy sets tackling the current status according to \textit{Nash Existence Theorem} and \textit{Bayesian Network Maximum A Posterior (MAP) Inference} \cite{koller2009probabilistic}. So \textit{GUT} also can be regarded as a \textit{Task-Oriented Decision Tree}. We formalize it as Theorem \ref{gut_dec} and Corollary \ref{gut_dec_map}. The detailed proof is given in the supplemental material (Appendix. \ref{proof}).

\begin{figure}[tbp]
\centering
\includegraphics[width=0.98\columnwidth]{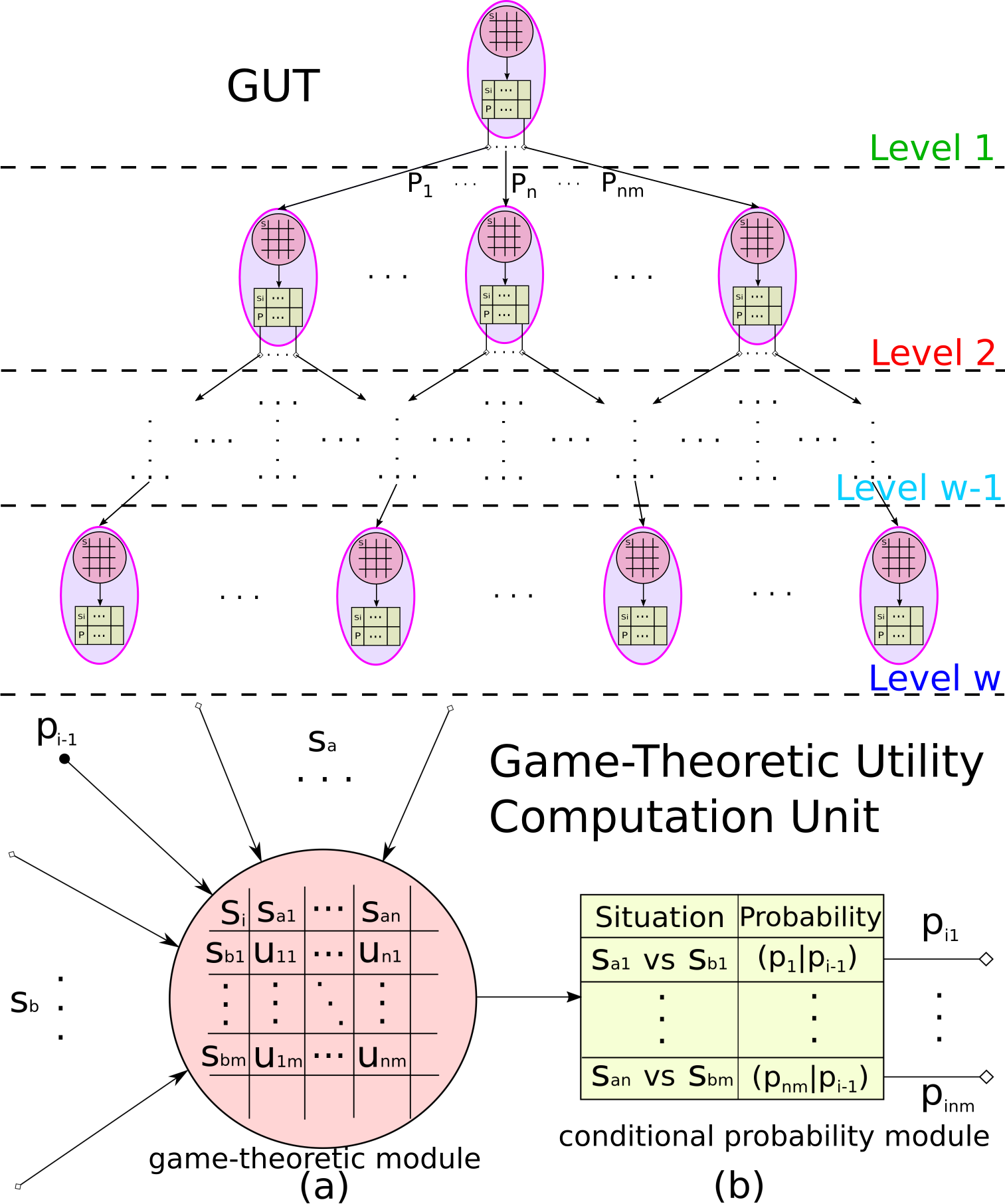}
\vspace{-2mm}
\caption{General Individual Agent's \textit{GUT}}
\label{fig:game_decision_trees}
\vspace{-4mm}
\end{figure}

\newtheorem{myTheo}{Theorem}
\begin{myTheo}[GUT Decision]
\label{gut_dec}
Let A and B represent the groups of Explorers and Aliens. The simultaneous normal-form game representing the non-cooperative game between explorers and aliens is a structure G=$\langle\{A,B\},\{S_e,S_a\},\{N_{t_A},N_{c_B}\}\rangle$. Supposing the GUT at the explorer group has w levels. G$_i\langle\{$A,B$\}$,$\{S_e,S_a\}$,N$_{t_{A_i}}\rangle$, i $\in$ w (Fig.~\ref{fig:game_decision_trees}.GUT) describes corresponding zero-sum game in each level. Then, A has at least one dominant strategy series (s$_1$, s$_2$, ... , s$_w$) in GUT.
\end{myTheo}

\newtheorem{myCoro}{Corollary}
\begin{myCoro}[GUT MAP]
\label{gut_dec_map}
Supposing the joint probability of solving a GUT is P(x) = P(x$_1$, x$_2$, ... , x$_w$). Assume we have a set of (exact or approximate) max-marginals $\{$MaxMarg$_P(X_i)\}_{X_i \in \chi}$ in all of the computation units $\chi$. Then, for each variable X$_i$(selected computation unit), there is a unique x$_i^*$ that maximize:
\begin{equation}
  x_i^* = \mathop{\arg\max}_{x_i \in S(X_i)} \mathop{MaxMarg_P(x_i)}
\label{gut_map_2}
\end{equation}
\end{myCoro}

\begin{table*}[t]
\begin{minipage}[b]{0.4\linewidth}
% \centering
    \begin{center}
    \setlength{\abovecaptionskip}{3pt}
    \scalebox{0.9}{
    \begin{tabular}{|c|c|c|}
    \hline
    \diagbox{AT}{Utility}{ET} & Attack & Defend \\ \hline
    Attack & $W_{AA}$ & $W_{DA}$ \\ \hline
 Defend & $W_{AD}$ & $W_{DD}$ \\ \hline
 \end{tabular}}\vspace{0mm}
 \caption{\small{Level 1 (Attack/Defend)- Explorer \& Alien Tactics Payoff Matrix.}}
    \label{first_level_matrix}  
    \setlength{\abovecaptionskip}{3pt}
    \scalebox{0.9}{
    \begin{tabular}{|c|c|c|c|}
    \hline
    \diagbox{AT}{Utility}{ET} & Nearest & A Lowest & A Highest \\ \hline
    Nearest & $E(e)_{NN}$ & $E(e)_{A_LN}$ & $E(e)_{A_HN}$ \\ \hline
 A Lowest & $E(e)_{NA_L}$ & $E(e)_{A_LA_L}$ & $E(e)_{A_HA_L}$ \\ 
 \hline
 A Highest & $E(e)_{NA_H}$ & $E(e)_{A_LA_H}$ & $E(e)_{A_HA_H}$) \\ 
 \hline
 \end{tabular}}\vspace{0mm}
    \caption{\small{Level 2 (Who to Attack/Defend)- Explorer \& Alien Tactics Payoff Matrix.}}
    \label{second_level_matrix}
    \scalebox{0.9}{
    \begin{tabular}{|c|c|c|c|}
    \hline
    \diagbox{AT}{Utility}{ET} & One Group & Two Group & Three Group \\ \hline
    Independent & $E(hp)_{1I}$ & $E(hp)_{2I}$ & $E(hp)_{3I}$ \\ \hline
 Dependent & $E(hp)_{1D}$ & $E(hp)_{2D}$ & $E(hp)_{3D}$ \\ \hline
 \end{tabular}}
    \caption{\small{Level 3 (How to Attack/Defend) - Explorer \& Alien Tactics Payoff Matrix.}}
    \label{third_level_matrix}
    \end{center}
\end{minipage}
\hspace{3mm}
\begin{minipage}[b]{0.55\linewidth}
\centering
\scalebox{0.9}{
\begingroup
\removelatexerror
\begin{algorithm}[H]
\small
\KwIn{Explorers' and Aliens' states. ($\beta$ = Aliens' number)}
\KwOut{formation shape $s$; current attacking target $t$; number of groups $g$.}
\caption{Explorer's Collective Strategy Using \textit{GUT} Model in \textit{Explorers and Aliens Game}.}
\label{alg:DP}
\BlankLine
set state = $"level~one"$; \\
\While{${\Delta|\beta|} >= 1 \; \textbf{And} \; |\beta|$ != 0 i.e., (at least one new alien)} 
{
    \uIf{state==$"level~one"$}
    {
        Compute the Nash Equilibrium; \\
        Get the most feasible formation shape $s$; \\
        state = $"level~two"$
    }
    \uElseIf{state==$"level~two"$ And s != Null}
    {
        Compute the Nash Equilibrium; \\
        Get the most feasible attacking target $t$; \\
        state = $"level~three"$
    }
    \ElseIf{state==$"level~three"$ And s, t != Null}
    {
        Compute the Nash Equilibrium; \\
        Get the most feasible number of groups $g$; \\
    }   
}

\If{$\beta$ == 0}
{
    $s$ = $"Patrol"$; \\
    $g$ = 1;
}
\Return $s,t,g$
\end{algorithm}
\endgroup}\vspace{3pt}
\end{minipage}
\vspace{0mm}
\end{table*}

\paragraph{Complexity Analysis}

Like \textit{the master theorem} \cite{cormen2009introduction}, supposing each sub-game has the same size (strategies space), the GUT can be described as the running time of an approach that recursively divides a game $G(\xi)$ of size $\xi$ into $a$ sub-games, each of size $\xi/b,~a,b \in Z^+$. If $G(\xi)$ is the one-level game, the complexity obeys \cite{daskalakis2009complexity}.
% the GUT describes the running time of an approach that recursively divides a game $G(n)$ of size $n$ into $a$ sub-games, each of size $n/b$, where $a$ and $b$ are positive constants. 
Then $G(\xi)$ has the following asymptotic bounds:
\begin{equation}
  \xi^{log_b^a} \leq G(\xi) \leq \xi^{log~\xi/\epsilon^2},~\epsilon \in (0,1).
\label{gut_map_2}
\end{equation}
It runs in $\mathop{O}(log_b^\xi)$ time on searching the specific strategy set, showing the scalability in the strategies space (game size). The scalability in terms of the number of agents depends on the particular communication graph in information sharing.

Alg.~\ref{alg:DP} demonstrates a three-levels GUT for the specific game in the "explore domain" (Fig.~\ref{fig:overview}). 
More specifically, the first level defines the agent's high-level strategies: \textit{Attack} and \textit{Defend}, which are represented as \textit{Triangle} and \textit{Regular~Polygon} formation shapes (Fig. \ref{fig:overview}) in the explorers. Based on the first level decision, they need to decide the specific opponent attacking or defending in the second level. Here, we assume that agents have three basic tactics: attacking or defending the \textit{nearest}, the \textit{lowest}, or \textit{highest} attacking ability adversary. In the last level, explorers choose to form how many groups and aliens select follow neighbors' behaviors or not. Table.~\ref{first_level_matrix}, \ref{second_level_matrix} and \ref{third_level_matrix} show the corresponding payoff matrix.

The \textit{Utility Function} design is critical to determine whether or not an individual can calculate reasonable tactics. In order to simplify the whole process, we adopt the winning probability ($W$), basic needs (energy cost  - $E(e)$), and safety needs (HP (health power) cost  - $E(hp)$) representing the expected utility values (See Appendix. \ref{needs_utility} for more details).

\subsection{Unintentional Adversary Decision-Making}

We design the \textit{Adapting The Edge} algorithm for the unintentional adversaries. It can help agents tackle static unintentional adversaries and adapt their edge's trajectory until it finds a suitable route to the goal point. Through sharing the communication data between agents, individuals can select the direction of less potential collision probability to move. In our scenarios, the two mountains represent the unintentional adversaries, and explorers need to find a path passing through them (See Appendix~\ref{unintentional_adversarial_decision} for more details).

\subsection{Explore Domain Implementation}

In our game, the explorers group as \textit{Patrol} formation (see Fig. \ref{fig:overview}) detecting the unknown world. After tackling various threats and adversaries, they always choose the shortest path to the goal point, then circle the treasure. In the whole process, explorers present a kind of global behaviors performing \textit{Collective Rationality} and caring about \textit{Group interest}. In contrast, aliens show \textit{Self-interest} and do not cooperate (See Appendix. \ref{relative_definition} for relative definitions). For explorers, their \textit{Teaming Needs} (expected utilities) is under the premise of maximizing the chance of finding the treasure to minimize HP cost based on fitting their low-level needs, such as safety and basic needs.
% \vspace{-3mm}

\begin{figure*}[t]
\centering
    \subfigure[Explorer Average HP Cost]{
    \begin{minipage}[t]{0.32\linewidth}
 \includegraphics[width=1\textwidth]{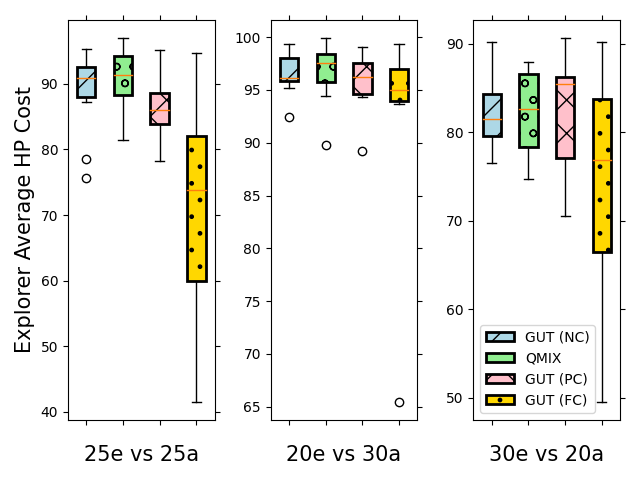}
 \label{fig: eahc}
 \end{minipage}}
 \subfigure[Kill Per Alien - Explorer Lost]{
    \begin{minipage}[t]{0.32\linewidth}
 \includegraphics[width=1\textwidth]{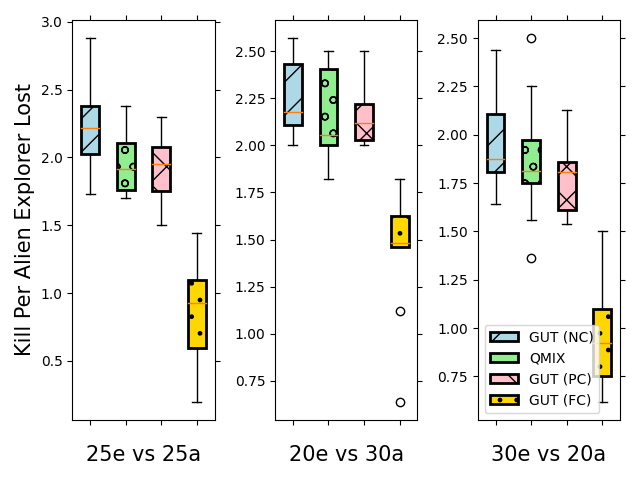}
 \label{fig: kpmel}
 \end{minipage}}
 \subfigure[Kill Per Alien - HP Cost]{
    \begin{minipage}[t]{0.32\linewidth}
    \includegraphics[width=1\textwidth]{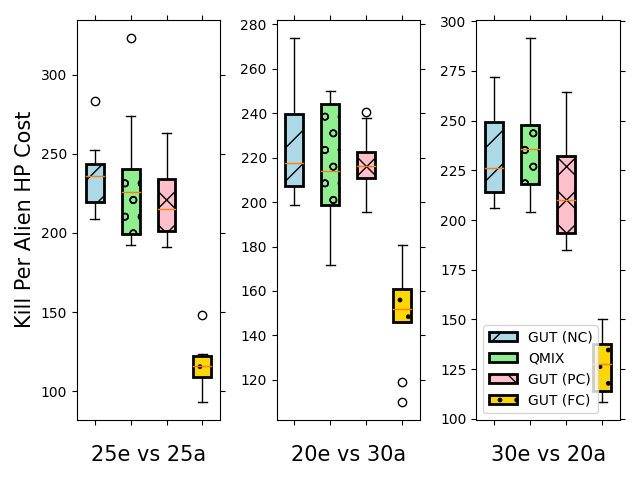}
    \label{fig: kpmhc}
    \end{minipage}}
    \vspace{-2mm}
\caption{\small{The Performance of Interaction Experiments with Different Proportion in Corresponding Scenarios.}}
\label{fig: interactive_experiment}
\vspace{-2mm}
\end{figure*}

%%%%%%%%%%%%%%%%%%%%%%%%%%%%%%%%%%%%%%%%%%%%%%%%%%%%%%%%%%%%%%%%%%%%%%%%
%\newpage
\section{Experiments}
\label{sec:evaluation}
% \vspace{-2mm}

We evaluate \textit{GUT} from two different perspectives: \textit{Interaction Experiments} compares the performance of explorers' cooperative strategies between \textit{GUT} and \textit{QMIX}; \textit{Information Prediction} demonstrates the \textit{GUT} when different predictive models are implemented to estimate aliens' states.

In experiments, we suppose each explorer has the same energy and HP levels initially, and every moving step will cost $0.015\%$ energy. Every communication round and per time attacking will cost $0.006\%$ and $0.01\%$ energy, respectively. Aliens have 3x more capable than explorers in the attacks, and per time attacking will cost the explorers $0.15\%$ HP. Their per time attacking energy and per time attacked HP cost are $0.03\%$ and $0.05\%$ (Appendix. \ref{experiment_setting} shows more details about the experiment setting).
The video demonstrating the experiments is available through an anonymous video hosting service at \url{https://streamable.com/gty9am}.

\begin{table*}[t]
% \centering
\begin{minipage}[b]{0.4\linewidth}
\centering
\setlength{\abovecaptionskip}{6.5pt}
\setlength{\belowcaptionskip}{1pt}
\includegraphics[width=1\columnwidth]{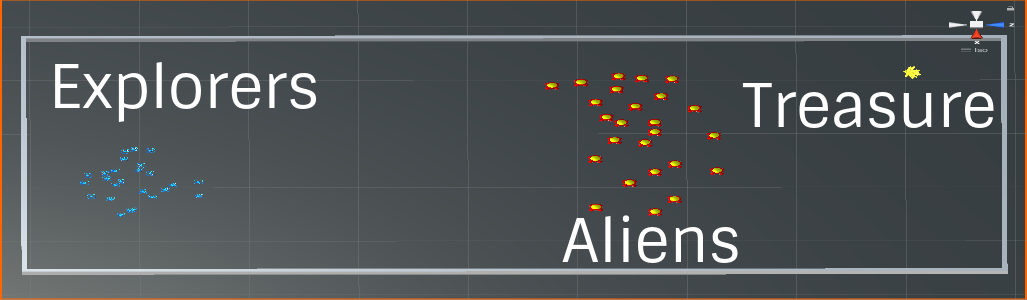}\vspace{-1.5pt}
\figcaption{\small{Only Intentional Adversary.}}
\label{fig: no_obstacles}

\scalebox{0.65}
{
\renewcommand\arraystretch{1.5}
\begin{tabular}{|c|c|c|c|c|}
    \hline
    \diagbox{PRD}{Winning Rate}{APP} & GUT (NC) & QMIX & GUT (PC) & GUT (FC)\\ \hline
    20e vs 30m & 40\% & 50\% & 50\% & 70\% \\ 
    \hline
 25e vs 25m & 90\% & 100\% & 100\% & 100\% \\ 
 \hline
 30e vs 20m & 100\% & 100\% & 100\% & 100\% \\ 
 \hline
 \end{tabular}}\vspace{5pt} 
 \tabcaption{\small{Winning Rate Comparison.}}
    \label{tab: wrc}
\end{minipage}
\hspace{1mm}
\begin{minipage}[b]{0.55\linewidth}
\setlength{\abovecaptionskip}{8pt}
\begin{center}
\scalebox{0.7}
{
\renewcommand\arraystretch{1.5}
\begin{tabular}{c|c|c|c|c|c|c|c|c|c}
\hline 
\multirow{2}*{Ra} &  \multicolumn{3}{c|}{Com} &  \multicolumn{3}{c|}{Incom L} & \multicolumn{3}{c}{Incom Poly}\\
\cline{2-10}
~&WR&C$_{s_e/w}$&C$_{s_{hp}/w}$&WR&C$_{s_e/w}$&C$_{s_{hp}/w}$&WR&C$_{s_e/w}$&C$_{s_{hp}/w}$\\
\hline
\multicolumn{10}{c}{With Intentional Adversary} \\
\hline
20:30&70\%&1077.37&2649.80&30\%&2367.14&6306.90&30\%&2726.64&6216.44\\
\hline
20:25&90\%&818.63&2027.98&50\%&1414.84&3807.23&40\%&1375.83&4824.64\\
\hline
25:25&100\%&1211.09&1772.00&90\%&1432.06&2606.12&80\%&1789.15&2949.89\\
\hline
25:20&100\%&1414.35&1739.78&100\%&1449.07&1960.45&100\%&1472.46&2177.52\\
\hline
30:20&100\%&1608.18&2241.09&100\%&2041.85&2370.76&100\%&1961.86&2271.48\\
\hline
\multicolumn{10}{c}{With Unintentional Adversary and Intentional Adversary} \\
\hline
25:25&100\%&1443.85&2110.91&70\%&2144.10&3143.63&60\%&2451.37&3742.98\\
\hline
\end{tabular}}\vspace{0pt}
\end{center}
\caption{\small{System Utility Comparison. Ra: Ratio of Explorers to Aliens,  WR: Winning Rate, C$_{s_e/w}$: system average energy cost winning a round, C$_{s_{hp}/w}$}: system average HP cost winning a round.}
\label{tab:Strategy_Comparision}
\end{minipage}
\end{table*}

\subsection{Interaction Experiments}
% \vspace{-2mm}

We analyze \textit{GUT} by simulating different cooperative styles and communication forms, comparing the performance with the state-of-the-art cooperative decision-making approach -- QMIX \cite{pmlr-v80-rashid18a} as follows:

% \vspace{-2mm}
\paragraph{1) GUT (NC)} [\textit{Noncooperation + No Communication}]
In this situation, explorers adopt \textit{GUT} computing the winning rate based on its perceiving information, but no communication, which means that it does not get the consistency to attack or defend the specific alien (Appendix.~\ref{experiment_setting} -- Fig.~\ref{fig: qmix_gut}).
%\vspace{-2mm}

% \vspace{-2mm}
\paragraph{2) QMIX} [\textit{Partial Cooperation + Partial Communication}]
QMIX \cite{pmlr-v80-rashid18a} is a state-of-the-art value-based method applied to reinforcement learning in MAS. Here, we only focus on the decision making part of QMIX, which considers the global benefit yielding the same result as a set of individual rewards. It allows each agent to participate in a decentralized execution solely by choosing greedy actions for its rewards. Accordingly, we assume that each explorer can cooperate, communicate, and share information with its observing explorers. Then through calculating the corresponding winning rate based on the number of its observing explorers and aliens, it chooses attacking or defending the specific \textit{hp lowest} target (Appendix.~\ref{experiment_setting} -- Fig.~\ref{fig: qmix}).
%\vspace{-2mm}

% \vspace{-2mm}
\paragraph{3) GUT (PC)} [\textit{Partial Cooperation + Partial Communication}]
Here, we consider the same situation as \textit{QMIX}, but explorers calculate the winning rate with \textit{GUT} and get the consistency to attack or defend the \textit{hp lowest} alien through partial communication (Appendix.~\ref{experiment_setting} -- Fig.~\ref{fig: gut_pc}).
%\vspace{-2mm}

% \vspace{-2mm}
\paragraph{4) GUT (FC)} [\textit{Full Cooperation (collective Rationality) + Full Communication}]
Lastly, we assume each explorer working in a full communication mode and making decisions by the \textit{GUT}. It means that every group member can share its information and get consistency through negotiation in the distributed system (Appendix.~\ref{experiment_setting} -- Fig.~\ref{fig: gut_fc}).
%\vspace{-2mm}

In these experiments, we do not involve \textit{Unintentional Adversary}(obstacles) (Fig.~\ref{fig: no_obstacles}) and consider three different proportions (A/E) between aliens and explorers as follow: \textit{20 explorers vs 30 aliens}, \textit{25 explorers vs 25 aliens} and \textit{30 explorers vs 20 aliens}. We assume that an agent can detect opponents' current state in its perception range. For each scenario, we conduct ten simulation trials for each proportion with same environment setting. 
Fig.~\ref{fig: interactive_experiment} shows that \textit{GUT} (FC) has the best performance compared with other cases. The GUT (NC), QMIX, and GUT (PC) do not have much difference between \textit{explorer average HP cost} results in Fig. \ref{fig: eahc}, but in Fig.~\ref{fig: kpmel} \textit{num. of explorers lost for killing an alien} and Fig.~\ref{fig: kpmhc} \textit{HP cost for killing an alien}, the QMIX and GUT (PC) show some advantage comparing with GUT (NC). For the winning rate comparison, Table. \ref{tab: wrc} also reflects the similar results.

% \vspace{-1mm}
\paragraph{Results}
This experiment shows that cooperation conduces to decrease the costs and boost the winning rate for more challenging tasks. More importantly, GUT can help agents representing more complex group behaviors and strategies, such as forming various shapes and separating different groups adapting adversarial environments in MAS cooperation. It vastly improves system performance, adaptability, and robustness. Besides, communication plays an essential role in cooperation, such as solving conflicts and getting consistency through negotiation. In GUT (NC) and QMIX, agents only share local information about the number of observing agents for naive attacking or defending behaviors. However, GUT (FC) present more complex relationships between agents' cooperation by organizing global communication data.

\begin{figure*}[t]
\vspace{-6mm}
\centering
    \subfigure[Explorer Average HP Cost]{
    \begin{minipage}[t]{0.32\linewidth}
 \includegraphics[width=1\textwidth]{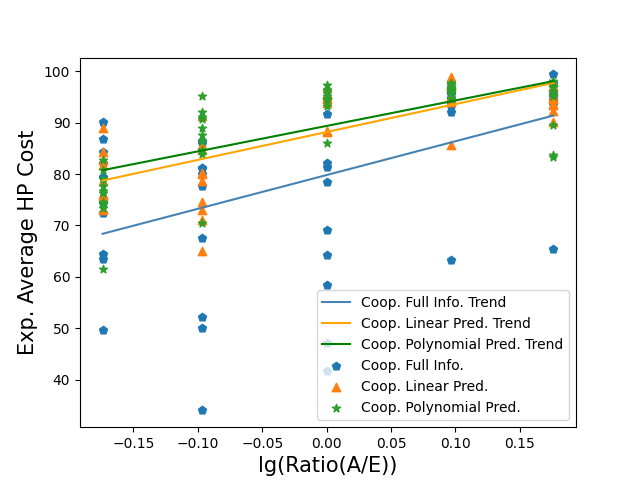}
 \label{fig: eahc_pre}
 \end{minipage}}
 \subfigure[Kill Per Alien - Explorer Lost]{
    \begin{minipage}[t]{0.32\linewidth}
 \includegraphics[width=1\textwidth]{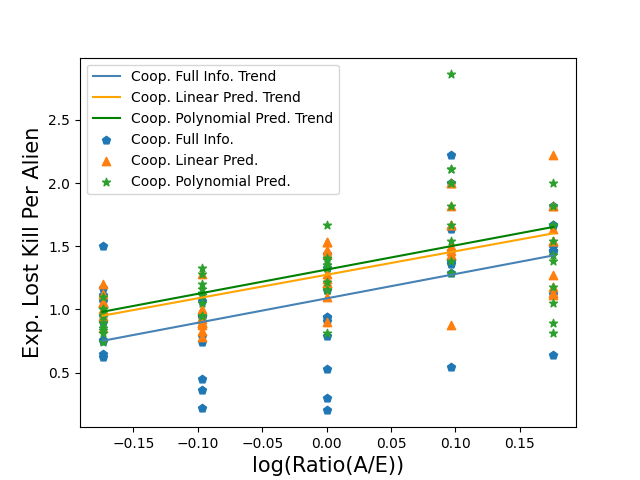}
 \label{fig: kpmel_pre}
 \end{minipage}}
 \subfigure[Kill Per Alien - HP Cost]{
    \begin{minipage}[t]{0.32\linewidth}
    \includegraphics[width=1\textwidth]{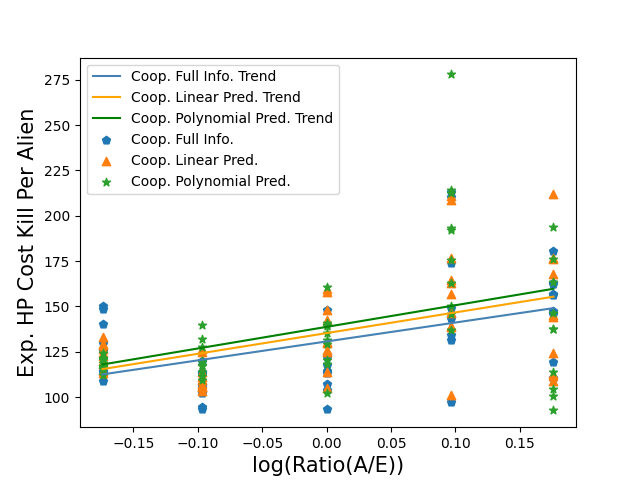}
    \label{fig: kpmhc_pre}
    \end{minipage}}
    \vspace{-4mm}
\caption{\small{The Individual Explorer's Performance with Different Predictive Models Only Intentional Adversary.}}
\label{fig: parameter_estimation_experiment}
\vspace{-4mm}
\end{figure*}

\begin{figure*}[t]
\centering
    \subfigure[\small{Explorer Average HP Cost/w}]{
    \begin{minipage}[t]{0.32\linewidth}
 \includegraphics[width=1\textwidth]{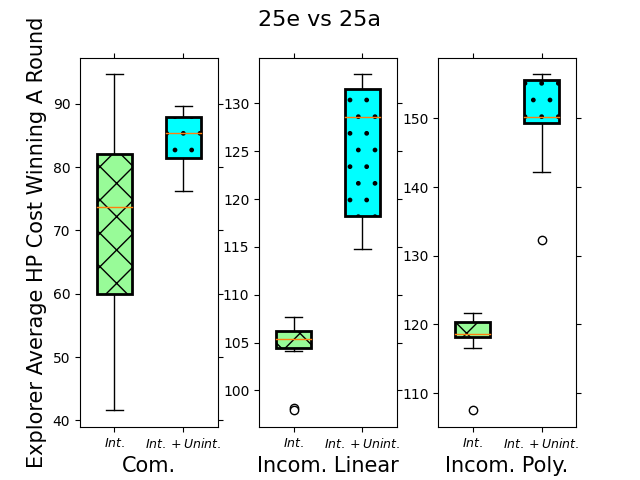}
 \label{fig: eahc_pre_mqmix}
 \end{minipage}}
 \subfigure[\small{Explorer Average Energy Cost/w}]{
    \begin{minipage}[t]{0.32\linewidth}
 \includegraphics[width=1\textwidth]{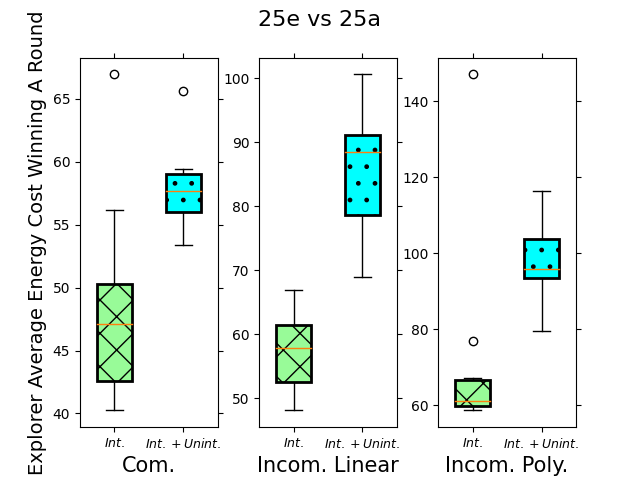}
 \label{fig: kpmel_pre_mqmix}
 \end{minipage}}
 \subfigure[\small{Explorer Average Loss/w}]{
    \begin{minipage}[t]{0.32\linewidth}
    \includegraphics[width=1\textwidth]{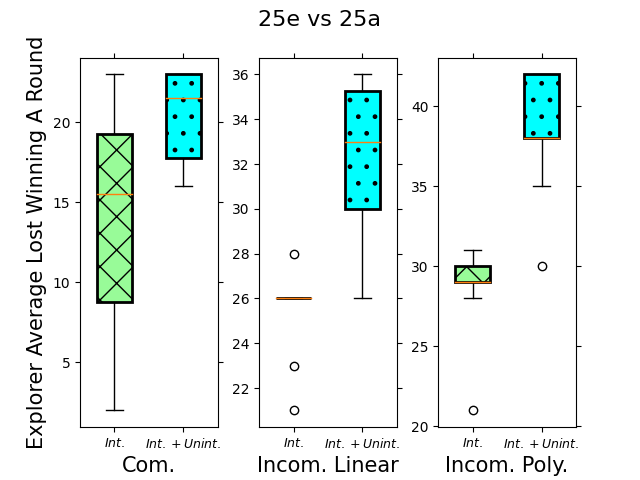}
    \label{fig: kpmhc_pre_mqmix}
    \end{minipage}}
    \vspace{-4mm}
\caption{\small{The Individual Performance with Different Predictive Models Considering Unintentional Adversary.}}
\label{fig: parameter_estimation_experiment_mqmix}
%\vspace{-4mm}
\end{figure*}

\subsection{Information Prediction}
\label{sec:prediction}
We design two kinds of perceiving models to analyze the individual and system performance in different scenarios. One is \textbf{\textit{Complete Information}}, which means that if an agent can perceive the adversary, it will detect the opponent's status, such as unit attacking energy cost and energy level. The other is \textbf{\textit{Incomplete Information}}. It implies that the agent can not gain opponents' state in its observable range.

%\vspace{-2mm}
\paragraph{Predictive Models}
We implement two \textit{Machine Learning} prediction models \textit{Linear} (Eq. \eqref{linear_regression}) and \textit{Polynomial Regression} (Eq. \eqref{polynomial_regression}), estimating adversaries' status in \textit{Incomplete Information}. We take regressors as individual unit cost \textit{HP$_{uc}$} and average system cost \textit{HP$_{asc}$} to predict opponent unit attacking energy cost $E_{uc}$ and current energy level $E_{el}$ respectively.
\begin{equation}
\begin{split}
    & E_{uc} = HP_{uc} \times \beta_{uc_0} + \varepsilon; \\
    & E_{el} = 100 - HP_{asc} \times \beta_{asc_0} + \varepsilon.
    \label{linear_regression}
\end{split}
\end{equation}
\begin{equation}
\begin{split}
    & E_{uc} = HP_{uc}^2 \times \beta_{uc_2} + HP_{uc} \times \beta_{uc_1} + \varepsilon; \\
    & E_{el} = 100 - HP_{asc}^2 \times \beta_{asc_2} -  HP_{asc} \times \beta_{asc_1} + \varepsilon.
    \label{polynomial_regression}
\end{split}
\end{equation}

Here, $\beta$ is corresponding regression coefficients($\beta_{uc_{0,1,2}}$ = \{0.08, 0.03, 0.0001\}, $\beta_{asc_{0,1,2}}$ = \{0.03, 0.0003, 0.00001\}), $\varepsilon$ presents the error following the normal distribution $\mathcal{N}(0,1)$.

%\vspace{-2mm}
\paragraph{1) With only intentional adversaries}

In this scenario, we consider five proportions of explorers and aliens (M/A) distributing in the map randomly. For each ratio, we also conduct ten simulation trials with the same experimental setting. 
From an individual perspective, Fig. \ref{fig: parameter_estimation_experiment} shows that \textit{Linear Regression} model has more accuracy than \textit{Polynomial Regression} model comparing with the result trend of \textit{Complete Information} (ground truth). From system perspective (Table.~\ref{tab:Strategy_Comparision}), the winning rate and system average energy/HP cost with different predictive models also show the similar results.

%\vspace{-4mm}
\paragraph{2) With intentional and unintentional adversaries}
In this setting, we consider a more complex scenario, which involves the unintentional adversary (two mountains) and aliens adopting the \textit{QMIX} to make their individual decision. We fix the number of explorers (E=25) and aliens (A=25) and conduct ten trails for each predictive model. Through the individual performance shown in Figs. \ref{fig: eahc_pre_mqmix} and \ref{fig: kpmel_pre_mqmix}, we notice that due to unintentional adversaries involved, individual average HP and energy cost for winning a round increase distinctly. Also, Fig. \ref{fig: kpmhc_pre_mqmix} shows that the entire group cost more agents to win a round concerning the obstacles involved. Table. \ref{tab:Strategy_Comparision} reveals  similar conclusion that unintentional adversaries lead to the decrease of the winning rate and more system cost with the same condition for winning one round. 

\paragraph{Results}
A suitable predictive model plays a vital role in shrinking biases or errors between the predictive results and ground truth through those experiments. More realistically, agents would face \textit{Incomplete Information} scenarios to estimate opponents' states from indirect information in adversarial environments. Furthermore, predictive models' parameters also require adapting corresponding scenarios, which means agents need to learn from their experience or system performance adjusting parameters for the specific situation.

\section{Conclusion and Future Work}
\label{sec:conclusions}
% \vspace{-2mm}

We introduce a new network model called \textit{Game-theoretic Utility Tree} (GUT) mimicking the agent decision-making process and the algorithm \textit{Adapting The Edge} for MAS cooperation working in adversarial environments. We then presented a new \textit{Explore Domain} evaluating GUT against the state-of-the-art cooperative decision-making approach QMIX. We demonstrated the effectiveness of GUT through two types of experiments including interaction and information prediction.

It will be essential for future work to improve GUT from different perspectives, such as optimizing GUT structure through learning from different scenarios, designing appropriate utility functions, building suitable predictive models, and estimating reasonable parameters fitting the specific scenario. Besides, implementing GUT in real robots is also an exciting and challenging problem helping us develop more robust computation models for MAS cooperation.

\clearpage

\bibliographystyle{named}
\bibliography{ijcai21}

\clearpage
\begin{appendix}
\setcounter{equation}{0}
\section*{Supplementary Material: \\ A Game-Theoretic Utility Network for Cooperative Multi-Agent Decisions in Adversarial Environments: }

\section{Relative Definitions}
\label{relative_definition}

In our experiments, we assume that the individual communication range is more extensive than its sensing range. Also, each agent can always connect at least one neighbor, which means that the entire group can build a connected communication graph in the whole process.

\paragraph{Self-interest (Individual Rationality)} Individual focuses on its own needs and desires  (interests). From philosophical, psychological, and economic perspectives, it is the motivation for intelligent agents to maximize their benefits or utilities from the individual perspective, which is also called \textit{Individual Rationality}.

\paragraph{Group-interest (Collective Rationality)} Comparing with \textit{Self-interest}, individual pays more attention about group members' needs, on the other hand, it is \textit{Group-interest}. Through negotiation, they finally get a consensus and agreement fulfilling a solution to maximizing the entire group's needs and rewards, which can be regarded as \textit{Collective Rationality}.

\paragraph{Partial Communication}
An individual agent only communicates and shares information with agents in the sensing or observable range.

\paragraph{Full Communication} 
In this status, agents always keep in touch with each other even when they are not in the sensing range. Also, the communication graph can be represented as a completely connected graph. It allows each agent to communicate and exchange data with its neighbor until the group reaches \textit{Information Equilibrium}, which means that every group member has the same information for the entire group \cite{9283249}.

\paragraph{Noncooperation} The individual does not communicate with each other and makes decisions only depending on its needs. In this situation, the agent only concerns its benefits based on \textit{Self-interest}.

\paragraph{Partial Cooperation} Based on the \textit{Partial Communication} information, individuals only cooperate with the observable group members to maximize their needs or minimize costs.

\paragraph{Full Cooperation} According to the \textit{Full Communication} data, individuals make decisions based on the \textit{Group-interest} showing \textit{Collective Rationality}.

\section{Experiment Setting} 
\label{experiment_setting}

\begin{figure*}[t]
\centering
% \setlength{\abovecaptionskip}{0pt}
% \setlength{\belowcaptionskip}{-10pt}
% \hspace{0mm}
\begin{minipage}[b]{0.245\linewidth}
\includegraphics[width=1\textwidth]{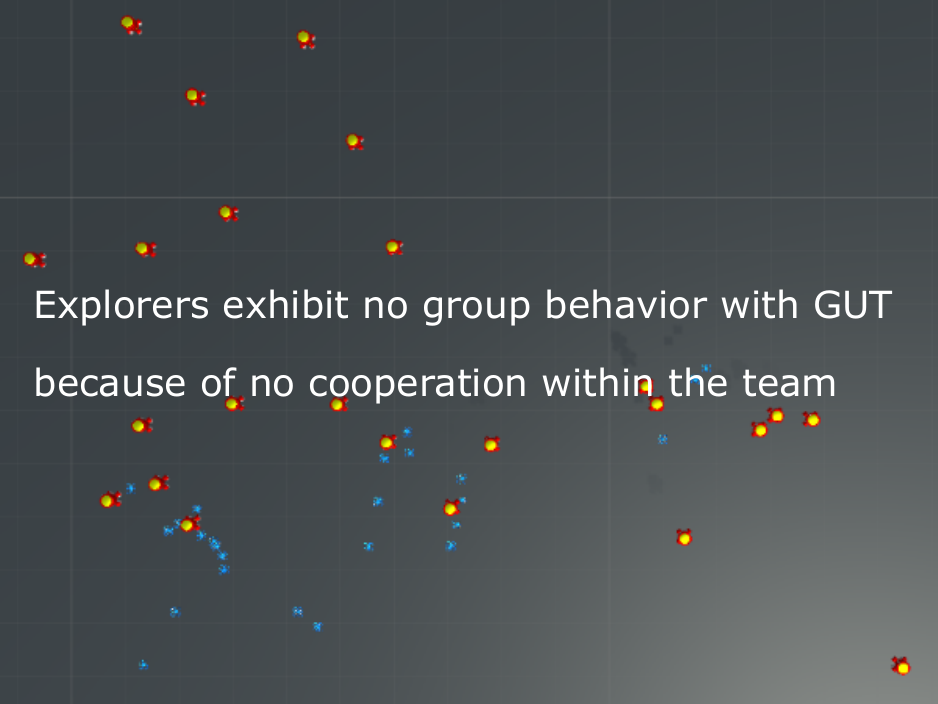}\vspace{0pt}
\caption{\small{GUT (NC)}}
\label{fig: qmix_gut}
\end{minipage}
\begin{minipage}[b]{0.245\linewidth}
\includegraphics[width=1\textwidth]{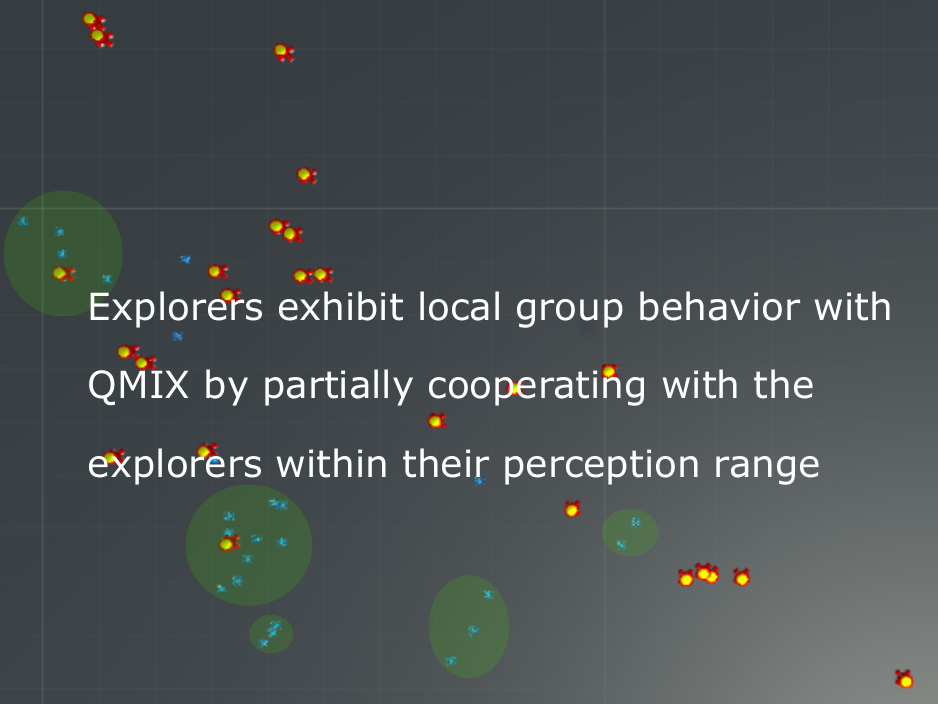}\vspace{0pt}
\caption{\small{QMIX}}
\label{fig: qmix}
\end{minipage}
\begin{minipage}[b]{0.245\linewidth}
\includegraphics[width=1\textwidth]{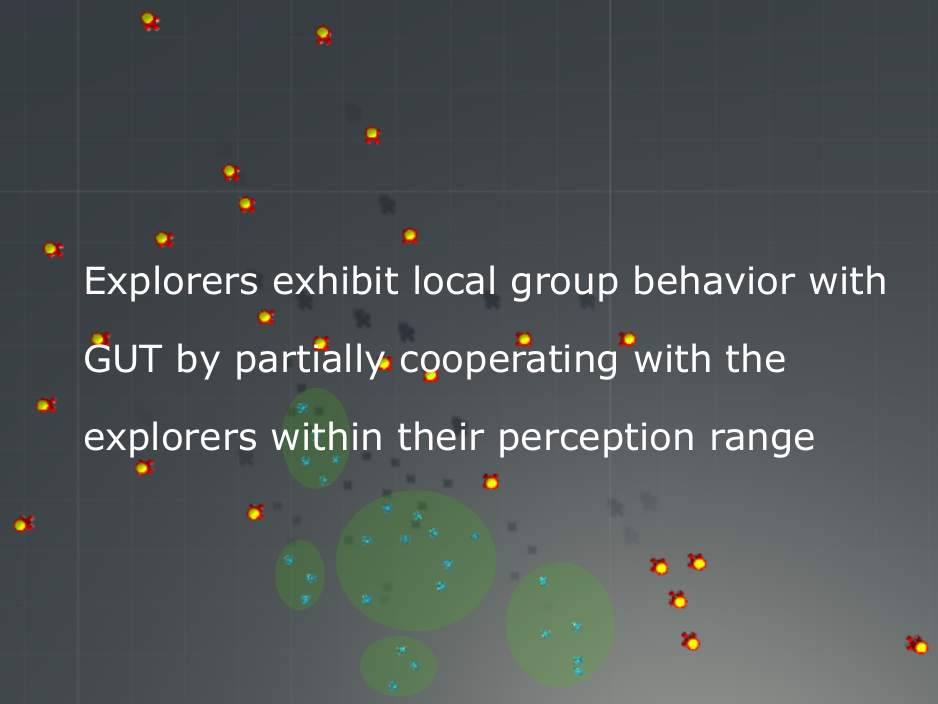}\vspace{0pt}
\caption{\small{GUT (PC)}}
\label{fig: gut_pc}
\end{minipage}
\begin{minipage}[b]{0.245\linewidth}
\includegraphics[width=1\textwidth]{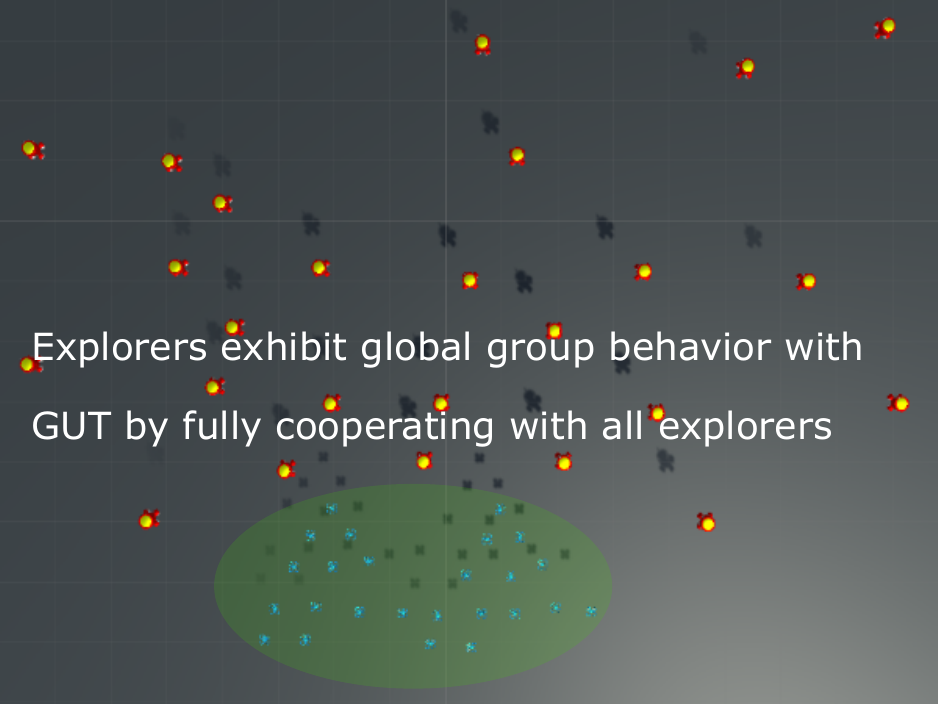}\vspace{0pt}
\caption{\small{GUT (FC)}}
\label{fig: gut_fc}
\end{minipage}
\vspace{-4mm}
\end{figure*}

Considering cross-platform, scalability, and efficiency of the simulations, we chose the ``Unity'' \cite{engine2008unity} game engine to simulate the \textit{Explorers and Aliens Game} and selected Gambit \cite{mckelvey2006gambit} toolkit
% \footnote{Gambit is an open-source collection of tools for doing computation in game theory which can build, analyze, and explore game models}
for calculating each level's Nash Equilibrium.

Each interaction (trial) in the \textit{Explorers and Aliens Game} last about 40 to 50 minutes in the simulated experiments on a laptop with Intel i7 Processor, GeForce GTX 1050 Ti GPU, and 16GB DDR4 RAM running the OS Ubuntu 18.04.

It is worth noting that an alien attacking capability is three times that of an explorer in all the experiments, which means that aliens represent higher capabilities (who act based on their self-interests) in preventing explorers from their tasks.

The video demonstration of the experiments showing sample trials of experiments using GUT and QMIX is available at the anonymized link \url{https://streamable.com/gty9am}.

\section{Unintentional Adversaries Decision}
\label{unintentional_adversarial_decision}

\begin{figure*}[t]
% \centering
\begin{minipage}[b]{0.42\linewidth}
\begin{center}
\includegraphics[width=0.9\columnwidth]{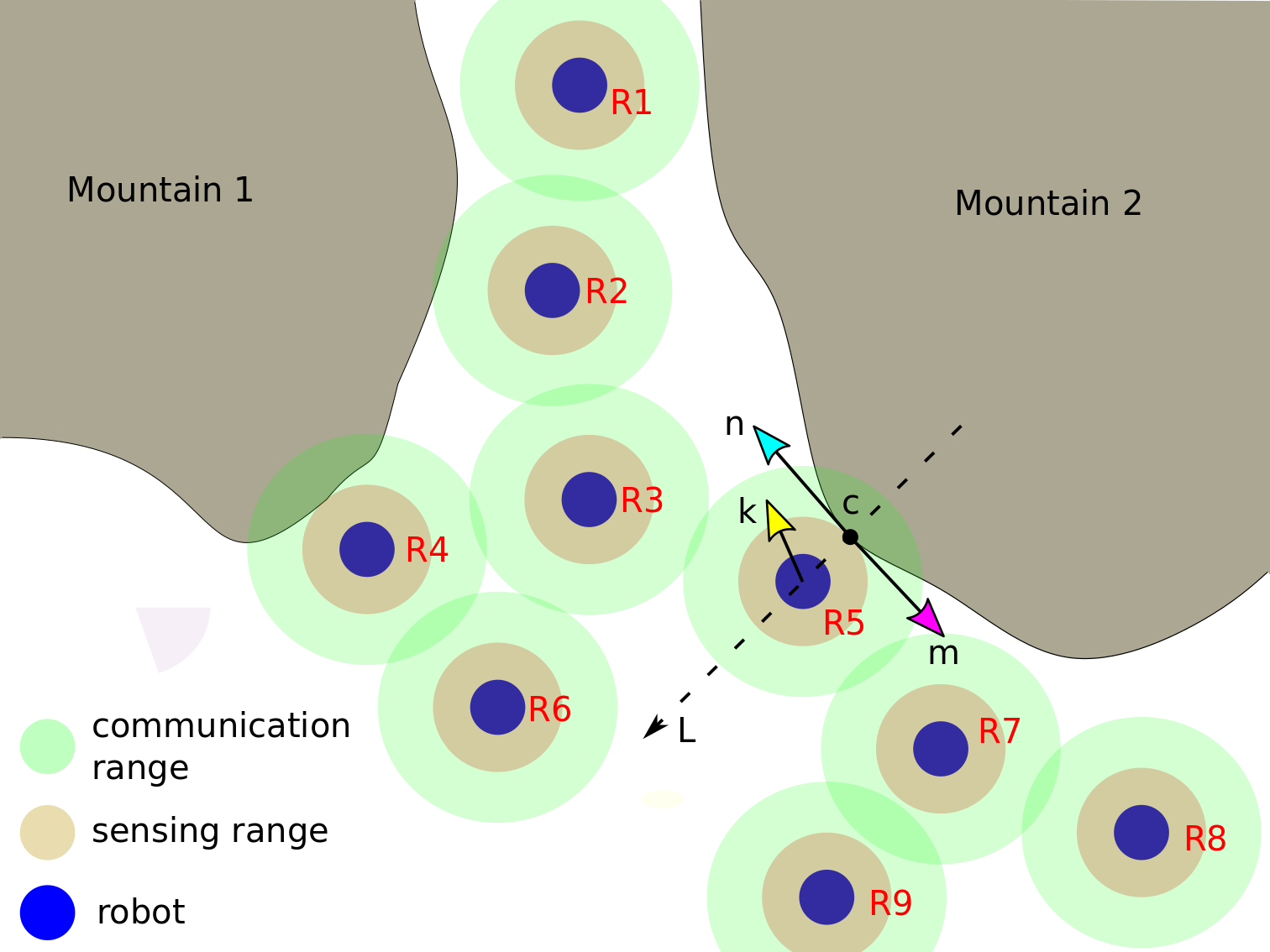}\vspace{-6pt}
\caption{\small{Illustration of "Adapt The Edge" algorithm for tackling (unintentional) obstacles. }}
\label{fig: fte}
% \vspace{-6mm}
\end{center}
\end{minipage}
\hspace{3mm}
\begin{minipage}[b]{0.52\linewidth}
\scalebox{0.95}{
\begingroup
\removelatexerror
\begin{algorithm}[H]
\small
\KwIn{Explorers' and Mountains' states}
\KwOut{moving direction $r$ and distance $\Delta d$}
\caption{Adapting The Edge}
\label{alg: fte}
\BlankLine
\While{The nearest collision point $c~!=Null$}
{
    calculate the number $n$ and $m$ of non-collision agents in both side of the line $l$ passing through $c$ and perpendicular $c$'s tangent;\\
    \uIf{$n > m$}
    {
        $r$ = $n$ side in line $l$; \\
        $\Delta d$ = one step of agent's movement;
    }
    \uElseIf{$n == m$}
    {
        agent stop;
    }
    \ElseIf{$n < m$}
    {
        $r$ = $m$ side in line $l$; \\
        $\Delta d$ = one step of agent's movement;
    }
}
\Return     $r$ = current position to goal point, $\Delta d$
\end{algorithm}
\endgroup}\vspace{0pt}
\end{minipage}
\end{figure*}

When explorers perceive the mountains (obstacles - static unintentional adversaries), they utilize limited information by sharing the perceiving information among agents for cooperative collision avoidance. Our experiments involve nine robots. In current situation (Fig.~\ref{fig: fte}), robots $R5$ detect the mountain. To avoid a collision, it needs to switch the moving direction $k$ based on the tangent's direction of the nearest collision point $c$. There are  two directions $n$ and $m$.  According to the current status, $R5$ will select the direction $n$, which has more non-collision robots potentially.

Specifically, $L$ is a straight line passing through the tangent point $c$ and perpendicular to $n$ and $m$. There are four robots $R1$, $R2$, $R3$, and $R6$ without collision in the direction $n$ comparing with three non-collision robots $R7$, $R8$, and $R9$ in the direction $m$ in current situation. So $R5$ will move $\Delta d$ in direction $n$, then adjust the direction to the goal point moving forward until not unintentional adversaries in its route through iterating the process. Alg. \ref{alg: fte} presents the decision process. 

\section{Analysis and Proofs}
\label{proof}
\subsection{\textbf{Decision-making using GUT}}
% \vspace{-2mm}

\newtheorem{proTheo}{Theorem}
\begin{proTheo}[GUT Decision]
\label{gut_dec}
Let A and B represent the groups of Explorers and Aliens. The simultaneous normal-form game representing the non-cooperative game between explorers and aliens is a structure G=$\langle\{A,B\},\{S_e,S_a\},\{N_{t_A},N_{c_B}\}\rangle$. Supposing the GUT at the explorer group has w levels. G$_i\langle\{$A,B$\}$,$\{S_e,S_a\}$,N$_{t_{A_i}}\rangle$, i $\in$ w (Fig.~\ref{fig:game_decision_trees}.GUT) describes corresponding zero-sum game in each level. Then, A has at least one dominant strategy series (s$_1$, s$_2$, ... , s$_w$) in GUT.
\end{proTheo}
% \begin{proof}
% see Appendix. \ref{gut_decision}
% \end{proof}

\begin{proof}
\label{gut_decision}
For \textit{w}-level GUT, supposing game G$_i$ in level \textit{k}, the size of action space of group A (the number of agent A is \textit{z} ) and B are \textit{l$_i$} and \textit{m$_i$} correspondingly. For the intentional decision, the \textit{zero-sum} game G$_i$ can be described as Eq. \eqref{level_game}:
\begin{equation}
    G_i=\langle\{A,B\},\{S_e,S_a\},N_{t_{A_i}}\rangle,~~~i \in w;\label{level_game}
\end{equation}

Based on the teaming needs (Eq.~\eqref{teaming_need}) definition, group A's expected utilities \textit{N$_{t_{A_i}}$} can be presented as Eq. \eqref{need_expectation}.
\begin{equation}
    N_{t_{A_i}}= \sum_{i=1}^{z} \mathbb{E}_i(U)=(u_{gk})_{l_i \times m_i},~~~g \in l_i, k \in m_i, j \in z;\label{need_expectation}
\end{equation}

According to \textit{Nash Existence Theorem}, it guarantees the existence of a set of mixed strategies for finite, non-cooperative games of two or more players in which no player can improve his payoff by unilaterally changing strategy \cite{weisstein2002crc}. So every finite game has a \textit{Pure Strategy Nash Equilibrium} or a \textit{Mixed Strategy Nash Equilibrium}.
The process can be formalized as two steps:

\textit{a. Compute Pure Strategy Nash Equilibrium}

We can present agents' utility matrix as Eq. \eqref{A_level1_utility}:\\
\begin{equation}
\left[
\begin{matrix} \label{A_level1_utility}
u_{11} & u_{12} & \cdots & u_{1m_i} \\ 
u_{21} & u_{22} & \cdots & u_{2m_i} \\ 
\vdots & \vdots & \ddots & \vdots \\ 
u_{l_i1} & u_{l_i2} & \cdots & u_{l_im_i} \\ 
\end{matrix} 
\right]
\end{equation}

The row and column correspond to the utilities of agent $A$ and $B$ separately. We can compute the maximum and minimum values of the two lists separately by calculating each row's minimum value and each column's maximum value.
\begin{equation}
\min \limits_{1\leq k\leq m_i} \max  \limits_{1\leq g\leq l_i} u_{gk} = 
\max \limits_{1\leq g\leq l_i} \min  \limits_{1\leq k\leq m_i} u_{gk} \label{p_game_result}
\end{equation}

If the two value satisfy the Eq. \eqref{p_game_result}, we can get the game G$_i$ Pure Strategy Nash Equilibrium Eq. \eqref{psne}, and corresponding game value Eq. \eqref{level1_game_result}.
\begin{eqnarray}
&& PSNE=(A_{g^*},B_{k^*}); \label{psne} \\
&& V_{G_i}=u_{g^*k^*} \label{level1_game_result}
\end{eqnarray}

\textit{b. Compute Mixed Strategy Nash Equilibrium}

The tactics' probability of agent $A$ present as Eq. \eqref{A_level1_game_probability}.
\begin{equation}
\begin{split}
    & AX =(x_1,x_2,\ldots,x_{l_i});\label{A_level1_game_probability} \\
    & \sum_{g=1}^{l_i} x_g=1, x_g \geq 0,~g=1,2,\ldots,l_i.
\end{split}
\end{equation}

Similarly, we also can conclude agent $B$ tactics' probability as Eq. \eqref{B_level1_game_probability}.
\begin{equation}
\begin{split}
    & BY=(y_1,y_2,\ldots,y_{m_i});\label{B_level1_game_probability} \\
    & \sum_{k=1}^{m_i} y_k=1, y_k \geq 0,~k=1,2,\ldots,m_i.
\end{split}
\end{equation}

We define \textit{(X, Y)} as \textit{Mixed Situation} in certain status. Then, we can deduce the expected utility of agent $A$ and $B$ Eq. \eqref{A_expected_utility} and \eqref{B_expected_utility} respectively.
\begin{eqnarray}
&& {\mathop{\mathbb{E}}}_A(X,Y)=\sum_{g=1}^{l_i} \sum_{k=1}^{m_i} u_{gk} x_g y_k = \mathop{\mathbb{E}}(X,Y); \label{A_expected_utility}\\
&& {\mathop{\mathbb{E}}}_B(X,Y)=-\mathop{\mathbb{E}}(X,Y) \label{B_expected_utility}
\end{eqnarray}

In the Game G$_i\langle\{$A,B$\}$,$\{S_e,S_a\}$,N$_{t_{A_i}}\rangle$, if we get all the \textit{Mixed Tactics} of agent $A$ and $B$ as Eq. \eqref{A_tactics} and \eqref{B_tactics}, we can deduct the G$_i$'s \textit{Mixed Expansion} as Eq. \eqref{mixed_expansion}. Furthermore, if a tactic $(X^*, Y^*)$ satisfies Eq. \eqref{A_mne_condition} and \eqref{B_mne_condition}, we define the tactic is the optimal strategy (Eq. \eqref{m_game_result}) in current state.
\begin{eqnarray}
&& S^*_e={AX}; \label{A_tactics} \\
&& S^*_a={BX}; \label{B_tactics} \\
&& G_i^*=\{S^*_e,S^*_a; \mathbb{E}\}; \label{mixed_expansion} \\
&& \mathop{\mathbb{E}}(X^*,Y)\geq V_{s_A}, \forall~Y \in S^*_e; \label{A_mne_condition} \\
&& \mathop{\mathbb{E}}(X,Y^*)\leq V_{s_B}, \forall~X \in S^*_a; \label{B_mne_condition} \\
&& V_{S_e}=V_{G_i}=V_{S_a} \label{m_game_result}
\end{eqnarray}

As the above discussion, we express the GUT computation process as corresponding Probabilistic Graphical Models \cite{koller2009probabilistic} -- \textit{Bayesian Network}. 

Supposing each node is independent, the total number of nodes N and the current \textit{joint probability distribution} of the group $A$ in the GUT can be represented as Eq. \eqref{num_nodes} and \eqref{joint_p}.
\begin{equation}
\begin{split}
    N & = N_1 + N_2 + ... + N_{w} \\
      & = 1 + l_1 \times m_1 + (l_1 \times l_2) \times (m_1 \times m_2) + ... \\
      & + (l_1 \times l_1 \times ... \times l_{n-1}) \times (m_1 \times m_1 \times ... \times m_{w-1}) \label{num_nodes}
\end{split}
\end{equation}
\begin{equation}
\begin{split}
    \mathbb{P}(X) & = \mathbb{P}(X_1, X_1, ..., X_{N}) \\
         & = \mathbb{P}(X_1) \mathbb{P}(X_2|X_1) ... \mathbb{P}(X_N|X_1, X_2, ... , X_{N-1}) \\
        %  & = \mathbb{P}(X_1) P(X_2|X_1) ... \mathbb{P}(X_N|X_{N-(l_1 \times l_2 \times ... \times l_{n-1}) \times (m_1 \times m_2 \times ... \times m_{n-1})}) \\
         & = \prod_i \mathbb{P}_i(X_i|Par_G(X_i)), i \in N
         \label{joint_p}
\end{split}
\end{equation}

Since \textit{Nash Existence Theorem} guarantees that every game has at least one Nash equilibrium \cite{jiang2009tutorial}, we get Eq. \eqref{exist_p}.
\begin{equation}
    \mathbb{P}_i(X_i) \neq 0 ~~ \Longrightarrow ~ \prod_i \mathbb{P}_i(X_i|Par_G(X_i)) = \mathbb{P}(X) \neq 0 \label{exist_p}
\end{equation}

\paragraph{Low Bound} If each level Nash Equilibrium calculation in the GUT is the Pure Strategy Nash Equilibrium, the individual agent can obtain a unique tactic entering into the next level, which means the tactic's probability is equal to one (Eq. \eqref{pure_s}). We also can get corresponding dominant strategy series (s$_1$, s$_2$, ... , s$_n$) in GUT.
\begin{equation}
    \mathbb{P}_i(X_i) = 1 ~~ \Longrightarrow ~ \prod_i \mathbb{P}_i(X_i|Par_G(X_i)) = \mathbb{P}(X) \equiv 1  \label{pure_s}
\end{equation}
\end{proof}
\vspace{-4mm}

% Furthermore, we can guarantee an \textit{w}-levels \textit{GUT} having the Maximum A Posterior (MAP) as follow: 

\newtheorem{proCoro}{Corollary}
\begin{proCoro}[GUT MAP]
\label{gut_dec_map}
Supposing the joint probability of solving a GUT is P(x) = P(x$_1$, x$_2$, ... , x$_w$). Assume we have a set of (exact or approximate) max-marginals $\{$MaxMarg$_P(X_i)\}_{X_i \in \chi}$ in all of the computation units $\chi$. Then, for each variable X$_i$(selected computation unit), there is a unique x$_i^*$ that maximize:
\begin{equation}
  x_i^* = \mathop{\arg\max}_{x_i \in S(X_i)} \mathop{MaxMarg_P(x_i)}
\label{gut_map_2}
\end{equation}
\end{proCoro}

% \begin{proof}
% see Appendix. \ref{gut_decision_map}
% \end{proof}

\begin{proof}
We can simplify an \textit{w}-level \textit{GUT} as one link \textit{Bayesian Network} (Fig. \ref{bn_map}).

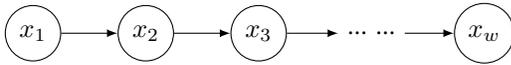
\begin{figure}[htpb]
\centering
\begin{tikzpicture}
[-latex,
node distance = 1cm]

\node[draw, circle] (n1) at (0,0)  {$x_1$};
\node[draw, circle] (n2) at (1.5,0)   {$x_2$};
\node[draw, circle] (n3) at (3,0)  {$x_3$};
\node (n4) at (4.5,0)  {... ...};
\node[draw, circle] (n5) at (6,0)  {$x_w$};
\draw (n1)--(n2);
\draw (n2)--(n3);
\draw (n3)--(n4);
\draw (n4)--(n5);

\end{tikzpicture}
\caption{w-level GUT as one link Bayesian Network.}
\label{bn_map}
\end{figure}

Now, we get the factors of product Eq.~\eqref{factor_p} ($\phi_{x_i}$ are the intermediate factors). So the maximum joint probability of \textit{GUT} Decision is equal to get its maximum factors of product. Then through VE (Variable Elimination) \cite{koller2009probabilistic}, we can get the MAP assignment of this \textit{GUT}. The entire process has two steps: 1) Variable elimination Eq. \eqref{ve}; 2) Tracing back to get a joint assignment ($x_1^*$, $x_2^*$, ... , $x_w^*$) Eq. \eqref{tb}. Finally, we can get MAP results of the \textit{GUT} Eq. \eqref{gut_map_1}.

\begin{equation}
\begin{split}
    \mathbb{P}(x) = & \prod_i \mathbb{P}_i(x_i|Par_G(x_i)) = \prod_{x_i} \phi_{x_i}~~ \Longrightarrow ~ \\
    & \max \mathbb{P}(x) = \max \prod_{x_i} \phi_{x_i}
\end{split}
\label{factor_p}
\end{equation}
\begin{equation}
\begin{split}
    & \textbf{Variable Elimination}: ~~~~~~~~~~~~~~~~~~~~~~~~~~~~~~~~~~~~~~~~~~~~~~~~~~~~~~~~~~~~~~~~~~~\\
    & \textbf{if}~X \in Scope[\phi_{x_i}]~~\textbf{then}~\mathop{\max}_X (\phi_{x_i}, \phi_{x_{i+1}}) = \phi_{x_i} \mathop{\max}_X (\phi_{x_{i+1}})~~ \\
    & \Longrightarrow\textit{first elimination}: \\
    & ~~~~~~\max \mathbb{P}(x) = \mathop{\max}_{x_2, x_3, ... , x_w} \phi_{x_2} \phi_{x_3} ...  \phi_{x_w} \mathop{\max}_{x_1} \phi_{x_1} \\
    & ~~~~~~\textit{second elimination}: \\
    & ~~~~~~\max \mathbb{P}(x) = \mathop{\max}_{x_3, x_4, ... , x_w} \phi_{x_3} \phi_{x_4} ...  \phi_{x_w} \mathop{\max}_{x_2} \phi_{x_2} \tau_1\\
    & ~~~~~~ ...~...\\
    & ~~~~~~\textit{(w)th elimination}: \\
    & ~~~~~~\max \mathbb{P}(x) = \mathop{\max}_{x_w} \tau_{w-1}\\
\end{split}    
\label{ve}
\end{equation}
\begin{equation}
\begin{split}
    & \textbf{Tracing Back}: ~~~~~~~~~~~~~~~~~~~~~~~~~~~~~~~~~~~~~~~~~~~~~~~~~~~~~~~~~~~~~~~~~~~~~~~~~~~~~~~~\\
    & ~~~~~~x_w^* = \mathop{\arg\max}_{x_w} \psi_w(x_w) \\
    & ~~~~~~x_{w-1}^* = \mathop{\arg\max}_{x_{w-1}} \psi_w(x_w, x_{w-1}) \\
    & ~~~~~~ ...~... \\
    & ~~~~~~x_1^* = \mathop{\arg\max}_{x_1} \psi_1(x_w, x_{w-1}, ... , x_1) \\
\end{split}    
\label{tb}
\end{equation}
\begin{equation}
\begin{split}
    & \textbf{Maximum A Posterior}: ~~~~~~~~~~~~~~~~~~~~~~~~~~~~~~~~~~~~~~~~~~~~~~~~~~~~~~~~~~~~~~~\\
    & ~~~~~~x^* = (x_1^*, x_2^*, ... , x_w^*)~\textit{is the MAP assignment}, \\
    & ~~~~~~\tau_{w-1}~\textit{is the probability of the most probable assignment.}
\end{split}    
\label{gut_map_1}
\end{equation}
\end{proof}

\section{The Definitions in Experiments}
\label{needs_utility}

In our experiments, we assume that explorers and aliens have the same moving speed, and aliens can not share information. Sec.~\ref{sec:terms} lists terms and notations used in this subsection. We implement the first two levels of agent needs hierarchy (safety needs - Health and basic needs - Energy). Capability needs and teaming needs are not implemented in this paper.

\subsection{Winning Utility Expectation}
We consider using \textit{Winning Probability} following \textit{Bernoulli Distribution} to represent individual high-level expected utility (teaming \& cooperation needs) in the first level (Eq. \eqref{winning_possibility}). 
\begin{equation}
   W(t_{ev},t_{mv},r_{ev},r_{mv},n,m)=(a_1\frac{a_2t_{ev}+a_3r_{ev}}{a_4t_{mv}+a_5r_{mv}})^{\frac{m}{n}}; \label{winning_possibility}
\end{equation}

\subsection{Energy Utility Expectation}
The second level's utility can be described as the relative \textit{Expected Energy Cost} (Eq. \eqref{expected_energy_cost}, \eqref{e_d}, \eqref{e_a} and \eqref{e_d}), which consists of three parts of energy costs: $moving$, $attacking$, and $communication$.
\begin{equation}
\begin{split}
    & E(d,v,f,q,n,m,\phi_e,\phi_m)= b_0 + b_1 \int_{-\infty}^{+\infty} (n - m) e_d(x) p_d(x,d) \mathrm{d}x \\
    & + b_{2} (\sum_{i=1}^{+\infty} ne_{a_e}(i,f)p_{a_m}(j,m \phi_m) -\sum_{j=1}^{+\infty} me_{a_m}(j,q)p_{a_e}(i,n \phi_e)) \\ 
                  & + b_3 \sum_{w=1}^{+\infty} ne_c(w) p_c(w,\frac{d}{v}); \label{expected_energy_cost}
\end{split}
\end{equation}
\begin{eqnarray}
&& e_d(x)= b_{11}x; \label{e_d} \\
&& e_a(x,y)= b_{12}xy; \label{e_a} \\
&& e_c(x)= b_{13}x \label{e_c}
\end{eqnarray}

In the attack-defend process, we also assume that individual action distance (Eq.~\eqref{p_d}), times of attacks (Eq.~\eqref{p_{a_e}}) and being attacked (Eq.~\eqref{p_{a_m}}), and the communication times (Eq.~\eqref{p_c}) follow \textit{Normal Distribution}, \textit{Poisson Distribution} correspondingly.
\begin{eqnarray}
&& p_d(x,d)= \frac{1}{\sqrt{2\pi}} e^{-\frac{(x-d)^2}{2}}; \label{p_d} \\
&& p_{a_e}(x,\lambda_e)= \frac{e^{-\lambda_e} \lambda_e^x}{x!}; \label{p_{a_e}} \\
&& p_{a_m}(x,\lambda_m)= \frac{e^{-\lambda_m} \lambda_m^x}{x!}; \label{p_{a_m}} \\
&& p_{a_m}(x,\frac{d}{v})= \frac{e^{-\frac{d}{v}} (\frac{d}{v})^x}{x!}; \label{p_c}
\end{eqnarray}

Then, we can simplify the Eq. \eqref{expected_energy_cost} as Eq. \eqref{s_expected_energy_cost}.
\begin{equation}
\begin{split}
E(d,v,f,q,n,m,\phi_e,\phi_m)= & b_0 + b_1 b_{11} (n - m) d + \\
                              & b_2 b_{12} n m (f \phi_m - q \phi_e) + b_3 b_{13} n \frac{d}{v};\label{s_expected_energy_cost}
\end{split}
\end{equation}

\subsection{HP Utility Expectation}
In the lowest level, we use the expected HP cost to describe the expected utility (Eq. \eqref{expected_HP_cost}, \eqref{h-1}, \eqref{h_e} and \eqref{h_m}).
\begin{equation}
\begin{split}
    & H(k,t_e,t_m,r_e,r_m,g,\phi_e,\phi_m)= \\
    & c_0 + c_{1} (\sum_{i=1}^{+\infty} k h(t_e,r_e,i)p_{h_m}(i,\phi_m)- \sum_{j=1}^{+\infty} gh(t_m,r_m,j)p_{h_e}(j,\phi_e)); \label{expected_HP_cost}
\end{split}
\end{equation}
\begin{equation}
    h(x,y,z) = \rho z (x+y) \label{h-1}
\end{equation}
\begin{eqnarray}
&& t_{e,m}(e_{e,m})= \gamma_{e,m} e_{e,m}; \label{h_e} \\
&& r_{e,m}(e_{e,m})= \delta_{e,m} e_{e,m}; \label{h_m}
\end{eqnarray}

$p_{h_e}(i,\phi_e)$ and $p_{h_m}(j,\phi_m)$ can be similarly described as formulas Eq. \eqref{p_{a_e}} and \eqref{p_{a_m}} correspondingly. Then, through simplifying the Eq. \eqref{expected_HP_cost}, we finally get Eq. \eqref{s_expected_HP_cost}.
\begin{equation}
\begin{split}
H(k,t_e,t_m,r_e,r_m,g,\phi_e,\phi_m)= & \\
                                      & c_0 + c_1 \rho [k \phi_m e_e (\gamma_e + \delta_e) - \\
                                      & g \phi_e e_m (\gamma_m + \delta_m)]; \label{s_expected_HP_cost}
\end{split}
\end{equation}

\subsection{Notations used in Sec.~\ref{needs_utility}}
\label{sec:terms}
\begin{itemize}
    \item $AT$ and $BT$ present the action space of group A and B correspondingly;
    \item $n$ and $m$ present the number of Explorers and Aliens respectively;
    \item $d$ presents the group average distance between two opponents;
    \item $v$ presents the agent's velocity;
    \item $i$ and $j$ present the times of attacks and being attacked;
    \item $w$ presents Explorers' communication times;
    \item $f$ and $q$ present the unit attacking energy cost of both sides agents respectively;
    \item $t_{ev}$ and $t_{mv}$ present average attacking ability levels of both sides respectively;
    \item $r_{ev}$ and $r_{mv}$ present average defending ability levels of both sides respectively;
    \item $t_e$ and $t_m$ present specific agent's attacking ability levels of both sides respectively;
    \item $r_e$ and $r_m$ present specific agent's defending ability levels of both sides respectively;
    \item $\phi_e$ and $\phi_m$ present individual agent's size;
    \item $k$ presents the number of Explorers' attacking simultaneously;
    \item $g$ presents the number of Aliens' attacking simultaneously;
    \item $a$, $b$, $c$, $\rho$, $\gamma$ and $\delta$ present corresponding coefficient;
    \item $e_e$ and $e_m$ present the current energy level of Explorer and Alien;
    \item $h$ presents the current $HP$ level of agent;
    \item $p$ presents the probability corresponding to the different section. 
\end{itemize}

\end{appendix}

\end{document}